\documentclass[aps,twocolumn,superscriptaddress,showpacs]{revtex4}
\usepackage{epsfig}
\usepackage{natbib}
\usepackage{amsmath}
\usepackage{times}
\usepackage{psfrag}
\usepackage{subfigure}
\usepackage{graphicx}
\usepackage{epstopdf}
\begin{document}
\title{Noise Effects on birhythmic  Josephson Junction  coupled to a Resonator}

\author{R. Yamapi}
\email{yamapi@yahoo.fr}
\affiliation{Department of Physics, Faculty of
 Science, University of Douala, Box 24 157 Douala, Cameroon.}

\author{G. Filatrella}
\email{filatrella@unisannio.it}
\affiliation{Department of Sciences and Technologies\\
and Salerno unit of CNISM, University of Sannio, Via Port'Arsa 11,
I-82100 Benevento, Italy.}

\date{\today}

\begin{abstract}
We study the effect of noise on a Josephson junction that, coupled to a linear $RLC$ resonator, can oscillate at two frequencies. To establish the global stability of the attractors, we estimate the position of the separatrix, 
{ an essential information to establish the stability of the attractor for this multidimensional system,}
from the { analysis} of the mean first passage time.  We find that the frequency locked to the resonator is most stable at low bias, and less stable at high bias, where the resonator exhibits the largest oscillations. The change in the birhythmic region is dramatic, for the effective barrier changes of an order of magnitude and the corresponding lifetime of about seven decades.

\end{abstract}

\maketitle

\newpage

\section{Introduction }
\begin{figure}[htbp]
\includegraphics[angle=0,height=7cm, width=10cm]{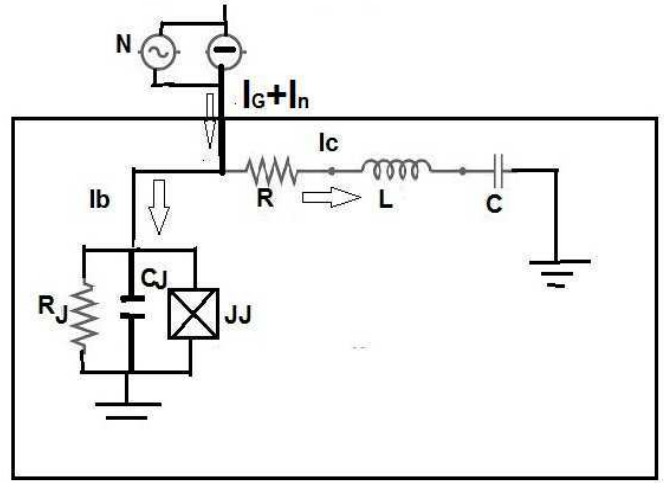}
\caption{\it Scheme of the Josephson Junction  coupled to a resonator. The current supplier  is at room temperature, while the JJ and the $RLC$ resonator are in the refrigerated box.}
\label{circuit}
\end{figure}

\begin{figure}[htbp]
\centerline{\includegraphics[scale = 0.5]{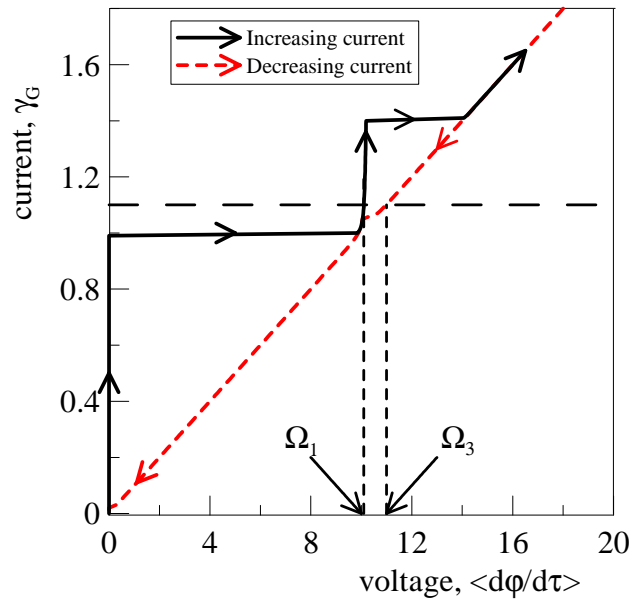}}
 \caption{\it (color online) Normalized $IV$ curve, for both increasing and decreasing current bias $\gamma_G$. Parameters of the simulation are: $\beta_l=0.01$, $Q_l=200$, $\Omega=2$, $\alpha=0.1$}
\label{IV}
\end{figure}

\noindent
The contemporary presence of two frequencies for the same set of parameters, or birhythmicity, is encountered in some biochemical systems \cite{decroly82,morita89,haberichter01,sosnovtseva02,abou-jaoude11}, nonlinear electronic circuits \cite{enjieu07,zakharova10,yamapi10,ghosh11,yamapi12,yue12}, and extended distributed systems \cite{stich02,casagrande05}. The experimental observation of birhythmic systems is, however, less frequent \cite{hounsgaard88,geva-zatorsky06,ventura07,gonzalez08}. In this context the superconducting circuit consisting of Josephson Junctions (JJ) coupled to a cavity \cite{hadley88,filatrella92,ozyuzer07}, as in Fig. \ref{circuit}, represents a preeminent example of birhythmic system that is also interesting for applications. 
{ 
The coupling among the junctions is supposed to be provided by a resonant cavity \cite{grib06,gross13,grib13}, thus when all the junctions are entrained it is essential to have a large current in the cavity, such that the junctions can be  entrained through the current in the resonator \cite{grib13}.
The state with a large current coexist with a state at lower power; the two states are clearly characterized by two different frequencies. This is the essential feature of
}
birhythmicity, the coexistence of two attractors characterized by two different amplitudes and frequencies: depending on the initial conditions, the system can produce
oscillations at two distinct periods.
Being the attractors locally stable, the system would however stay at a single frequency,
the one selected by the choice of the initial. Thus the system exhibits an hysteretic behavior: the displayed frequency depends upon the initial conditions. 
{ In the presence of noise the system can switch from an attractor to the other under the influence of the random term.}
Birhythmicity is therefore a nonlocal phenomenon that cannot be investigated by linear analysis \cite{landauer78}.
In this work we aim to determine the global stability of the two states at different frequencies $\Omega_1$ and $\Omega_3$ of the $IV$ on Fig. \ref{IV}, to ascertain the birhythmic properties induced by the $RLC$ circuit. 
From the simulated $IV$ of Fig. \ref{IV} it is evident that at the same bias point, e.g. $\gamma_G=1.1$, two frequencies appear , viz. $\Omega_1$ and $\Omega_3$, depending on the initial conditions. 
The first frequency is reached increasing the bias current from zero on the Josephson supercurrent, while the second is obtained decreasing the current from high values on the resistive McCumber branch; the selection of the frequency actually displayed is thus determined by the initial conditions. 
{ 
The features of the IV depend upon other factors such as the number of JJs and the features of the resonator \cite{grib06,grib13}. Also, heating effects are believed to be relevant for synchronization \cite{wang10}, as well as coupling through charge transfer through the Josephson channel \cite{ovchinnikov13}. In this work we consider the simplest case of a single JJ coupled to a high $Q$ cavity, and we neglect heating, that occurs at a much slower time scale.
}

A switch from an attractor to the other is of central interest for devices based on synchronization of JJ through an $RLC$ circuit, in particular for BSCCO stacks for THz generation \cite{tachiki11}.
{ Applicationwise, it is undesirable an uncontrolled switch from the state locked to the $RLC$ (the high power generation) to the other (the low power emission) \cite{grib13}. 
Unfortunately, the analysis of large fluctuations, as large as to carry the system from an attractor to another, is not easy, for it goes beyond the linear stability \cite{kautz94,lin11} given by Lyapunov exponents \cite{lin11,tsang91}.}
{ 
In equilibrium non dissipative systems, global stability is given by the time $\kappa$ to escape from the energy potential $\Delta U$ at a given a noise level $D$. Arrhenius law predicts that the average escape time  $\kappa$ exponentially depends upon the ratio between the energy barrier and the noise intensity \cite{hanggi90}: }

\begin{equation}
\kappa \propto \exp \left[ \frac{\Delta U}{D} \right].
\label{arrhenius}
\end{equation}

\noindent  In nonequilibrium systems, or when the potential energy is not available, a possibility is to reverse the logic and to define a pseudopotential energy barrier proportional to the logarithm of the lifetime \cite{dykman79,graham85,graham86,dykman94,dykman01} , viz. 

\begin{equation}
\Delta U \equiv \lim_{D\rightarrow 0}  \left[ D \log(\kappa) \right].
\label{pseudopot}
\end{equation}

\noindent  In fact, under general assumptions, it can be postulated that the escape time between the two attractors exponentially depends upon a quantity (the pseudopotential energy) and it is inversely proportional to the noise intensity \cite{dykman79,graham85,graham86,dykman01}. 
This approach has been used to determine the energy barrier of  vortex motion \cite{kautz94,girotti07} in Josephson systems, and has been employed to determine the  energy barrier for anharmonic oscillators with cubic \cite{dykman94} and quintic \cite{yamapi10,yamapi12} nonlinearities. 
{ 
The same methods has been also used to investigate Shapiro steps \cite{kautz88}.
At variance with  irradiated JJ where one frequency is given by an external drive, in the present system the system self-generates the two frequencies.
Moreover, chaos can occur in rf-fields \cite{kautz88}, as well as in several JJ coupled together \cite{kolachi13}.
Instead we prefer to focus on a simpler system, where the switch only occurs because of noise, between two otherwise (locally) stable attractors.}

{ If the energy barrier is to be determined by means of the lifetime, as per Eq.(\ref{pseudopot}), it is crucial to locate the separatrix between the two basins of attraction of the stable states. 
To determine the basins of attraction requires the knowledge of the initial conditions that lead to one or the other of the stable solutions, and therefore demands a detailed exploration of the phase space. }
{ However, being this exploration very difficult in the four-dimensional system of Fig. \ref{circuit}, we propose to exploit the fact that $\Delta U$ is a Lyapunov function \cite{graham85} to estimate the separatrix. 
As will be shown in Sect. IV, the method we propose is capable to determine an effective threshold for the escape time, and therefore our approach constitutes a method for the estimate of the pseudopotential when the boundary of the basin of attraction is not exactly known.
With this approach, we find that the stability of the attractor is not uniform: at the bottom of the step the trapping energy is high, and decreases at the top.
This behavior is somehow counterintuitive, in that the global stability is enhanced when the frequencies of the two attractors get closer.}

The work is organized as follows. In the next Section we describe an underdamped JJ coupled to a resonator and subject to external bias and noise. In Sect. III we discuss the locally stable attractors characterized by two frequencies, and how a transition from an attractor to the other can occur under the influence of noise. In Sect. IV we describe the method to locate the separatrix between the two attractors, an essential information to reconstruct the activation barrier. The methodological premises permit to determine the stability properties of the JJ in the birhythmic region. Section V concludes.

\section{Model of a Josephson junction coupled to a resonator}

Figure \ref{circuit} schematically describes the model used in our analysis:
an underdamped JJ connected in parallel to an $RLC$ resonator. 
Both elements are supposed in the temperature controlled vessel, while the bias current is supplied by a device at room temperature. 
\begin{figure}
\centerline{\includegraphics[angle=0,height=8cm, width=8cm]{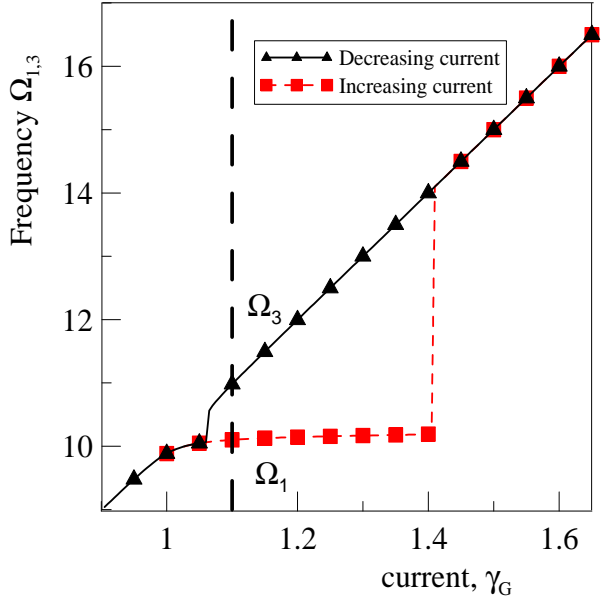}}
 \caption{\it (color online) Frequency $\Omega$ of the oscillations of the Josephson phase as a function of the bias current $\gamma_G$ for the deterministic case, $D=0$. Squares refer to increasing bias, and triangles to a decreasing bias, see Fig. \ref{IV}. The dashed line is the bias point of Fig. \ref{attractor}. Parameters of the simulation are: $\beta_l=0.01$, $Q_l=200$, $\Omega=2$, $\alpha=0.1$}
\label{omega}
\end{figure}
In this configuration the noise from the bias supply dominates respect to the Johnson noise from the resistors $R_j$ and $R$. Alternatively, one could add a random term for each resistor, as done for instance in Ref. \cite{lin11}. 
However, the noise is but a tool. Our goal is to determine the pseudoenergy; the principle of minimum energy \cite{graham86,kautz88} assures that the contributions from the minimal trajectory determines the height of the trapping potential, and therefore one does not expect substantial changes with a different noise source.

The electrical model consists of the capacitor $C_J$, the resistor $R_J$, and the ideal Josephson element, connected in parallel. The nonlinear relation between the current and the gauge invariant phase difference $\phi =\phi_1-\phi_2$ across two superconductors:
\begin{eqnarray}
\label{eqJcurrent}
I_J=I_0\sin \phi;
\end{eqnarray}

\noindent together with the Josephson voltage relationship
\begin{eqnarray}
\label{eqJvolt}
V_J=\frac{\hbar}{2e}\frac{d\phi}{dt}
\end{eqnarray}

\noindent determines that a JJ is an active oscillator that converts a dc current into an ac drive for the $RLC$ resonator. To derive the equations governing the system, we indicate with $I_C$ the current flowing through the $RLC$ circuit and with $\tilde q$ the charge on the capacitor. The JJ and the resonator are both biased by a current generator  $I_G$ affected by a noise current $I_n$ that split in the current $I_b$ through the JJ element and the current $I_C$ through the $RLC$. If we indicate with $I_{R_J}$ the current through the JJ resistor and with $I_{C_J}$ the current through the junction capacitance, we obtain the current balance:

\begin{equation}
\label{currbalance}
I_b=I_G+I_n-I_C
\Longrightarrow I_J+I_{R_J}+I_{C_J}=I_G+I_n-I_C.
\end{equation}

\noindent The Kirchhoff law for the loop voltage 

\begin{eqnarray}
\label{voltRLC}
V_J=V_C+V_{R}+V_L
\end{eqnarray}

\noindent completes the model, that is thus described by two second order coupled differential 
equations:

\begin{eqnarray}
\label{eqJJ_RLC}
\left\{\begin{array}{l}
\frac{C_J\hbar}{2e}\frac{d^2\phi}{dt^2}+\frac{\hbar}{R_J 2e}\frac{d\phi}{dt}+I_0\sin\phi+\frac{d\tilde q}{dt} = I_G+I_n , \\
\frac{d^2\tilde q}{dt^2}+\frac{R}{L}\frac{d\tilde q}{dt}+\frac{1}{LC}\tilde q-\frac{\hbar}{2eL}\frac{d\phi}{dt}  =   0 .
\end{array}\right.
\end{eqnarray}

\noindent Introducing the Josephson frequency $\omega_j=\sqrt{2eI_0 / C_J\hbar}$, Eqs.(\ref{eqJJ_RLC}) can be cast in the normalized units $\tau=\omega_j t$ and $q = \omega_j \tilde q /I_0$:

\begin{eqnarray}
\label{eqJJ_RLC_norm}
\left\{\begin{array}{l}
\frac{d^2\phi}{d\tau^2}+\alpha\frac{d\phi}{d\tau}+\sin\phi+\frac{d q}{d\tau}=\gamma_G+\zeta\\
\frac{d^2 q}{d\tau^2}+\frac{1}{Q}\frac{d q}{d\tau}+\Omega^2 q-\frac{1}{\beta_L}\frac{d\phi}{d\tau}=0
\end{array}
\right.
\end{eqnarray}
where

$$Q=\frac{L}{R}\sqrt{\frac{2eI_0}{C_J\hbar}}, 
\, \alpha=\frac{1}{R_J}\sqrt{\frac{\hbar}{2eI_0C_J}}, $$
$$\, \beta_L=\frac{2eLI_0}{\hbar}, 
\, \gamma_G=\frac{I_G}{I_0},
\, \Omega=\frac{1}{\omega_j\sqrt{LC}} .    $$

The statistical features of the noisy term $\zeta$ are determined by:
\begin{eqnarray}
\label{correl}
<\zeta(\tau)> &=& 0, \nonumber \\
<\zeta(\tau)\zeta(\tau')> &=& 4 D \delta(\tau-\tau').
\end{eqnarray}

\noindent The noise is due to an external source, therefore it does not obey the fluctuation-dissipation theorem and it is independent of the resistance. 
Equations (\ref{eqJJ_RLC_norm},\ref{correl}) are simulated with the Euler algorithm \cite{fox88}. 
{ 
Deterministic results have also been simulated with a Runge-Kutta algorithm. 
The $IV$ curves have been obtained slowly increasing the bias current, with a step $\Delta \gamma_G \simeq 0.01$, and using the final state at the previous current step as the initial state for the increased (or decreased) current bias. At each current step a transient of about $1000$ normalized time is discarded. The averages are also calculated over the same time. 
}
The time step $\Delta \tau$ is, through all simulations, $\Delta \tau=0.0001$ for the Euler algorithm, and $\Delta \tau = 0.01$ for the Runge-Kutta deterministic simulations.
The stochastic results are also averaged over as many realizations as it is necessary to have reliable results. Finally, the {\it Box-Mueller} algorithm \cite{knuth69} is used to generate Gaussian white noise from two random numbers $a$ and $b$ which are uniformly distributed on the unit interval $[0,1]$. Thus for each step $\Delta \tau$, $\zeta_{n}$ is randomly distributed as follows:
\begin{eqnarray}
\label{random}
a = \textrm{random  number},\quad b = \textrm{random  number}, \nonumber \\
\zeta_{n} = \sqrt{-4D\Delta \tau \log(a)}\cos(2\pi b)
\end{eqnarray}

The system (\ref{eqJJ_RLC_norm}), depending on the initial conditions,  can exhibit oscillations at two distinct periods. 
If the resonator and the JJ are weakly coupled ($1/\beta_L << 1$), the two frequencies $\Omega_1$ and $\Omega_3$ are substantially unperturbed and correspond to the resonant frequency $\Omega$ of the $RLC$ and the unperturbed ($1/\beta_L=0$) frequency of the junction, respectively \cite{lin11,filatrella03}. 
{ In contrast with previous studies \cite{filatrella92,filatrella03}, this is the strong coupled limit  ($1/\beta_L=100$), and therefore $\Omega_1 \neq \Omega$ \cite{shukrinov12a,shukrinov12b}, as shown in Fig. \ref{IV} by the normalized voltage $<d\phi/d\tau>$ in correspondence of the applied current $\gamma_G$. The shift in voltage due to the interaction has been estimated in the limit case of a non dissipative resonator \cite{shukrinov12a}; in our normalizations it reads:
\begin{equation}
\omega_{res}=\Omega\sqrt{1+\frac{1}{\beta_L}}.
\label{shift}
\end{equation}
The quantitative agreement is poor, as expected for a dissipative cavity. 
However, the pure $LC$ cavity correctly predicts the trend towards an increase of the resonant frequency.

A word about normalization. First, in these units the normalized voltage and the normalized frequencies are expressed in the same units; therefore (see Fig. \ref{IV}) the resonant frequency should be comparable with the characteristic frequency of the junction, thus:
$$
\frac{1}{\sqrt{LC}} \simeq \omega_j .
$$
\noindent Also, the resonator capacitance and the junction capacitance are connected by the relation:
$$
C = \frac{1}{\beta_L\Omega^2} C_j .
$$
For large coupling ($1/\beta_L>>1$) the capacitance of the resonator should much larger than the junction capacitance. Finally, the relation between the resistance of the resonator $R$ and the resistance of the junction $R_j$ reads:
$$
R = \frac{\alpha \beta_L}{Q} R_j,
$$
thus for high $Q$ ($Q=200$ in these simulations) and large coupling ($1/\beta_L=100$), we get $R \simeq \alpha R_j$. For underdamped junctions (here $\alpha = 0.1$) the resistance of the resonator is much less than the resistance of the junction. However, some care should be taken: in the equivalent circuit of Fig. \ref{circuit} the resistance of the JJ is in parallel, while the resonator is modeled by series lumped elements. 
}

\begin{figure}
\centerline{\includegraphics[angle=0,height=6cm, width=10cm]{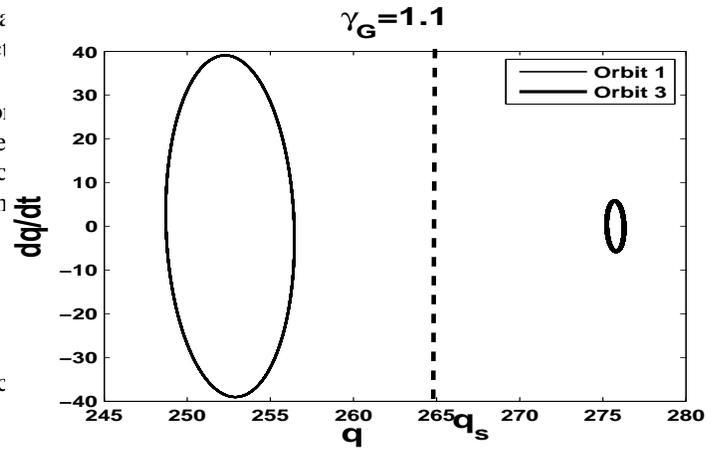}(i)}
\centerline{\includegraphics[angle=0,height=6cm, width=10cm]{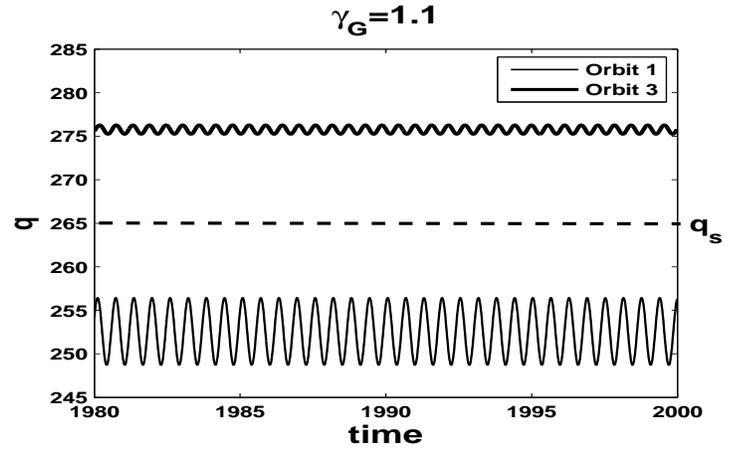}(ii)}
\caption{\it Projection of the phase space in the $q-dq/d\tau$ plane (i) and time evolutions (ii) for the deterministic dynamics, $D=0$.
Parameters of the simulation are: $\beta_l=0.01$, $Q_l=200$, $\Omega=2$, $\alpha=0.1$.}
\label{attractor}
\end{figure}

The resonant step locked to the cavity is shown in more detail in Fig. \ref{omega}: In the range $1.05 < \gamma_G <1.40$ the system stays on one or the other frequency, depending on the initial conditions (that are controlled by the bias sweep). 
To each frequency corresponds a different attractor, as will be analyzed in the next Section.

\begin{figure}
\centerline{(i)}
\centerline{\includegraphics[angle=0,scale=0.5]{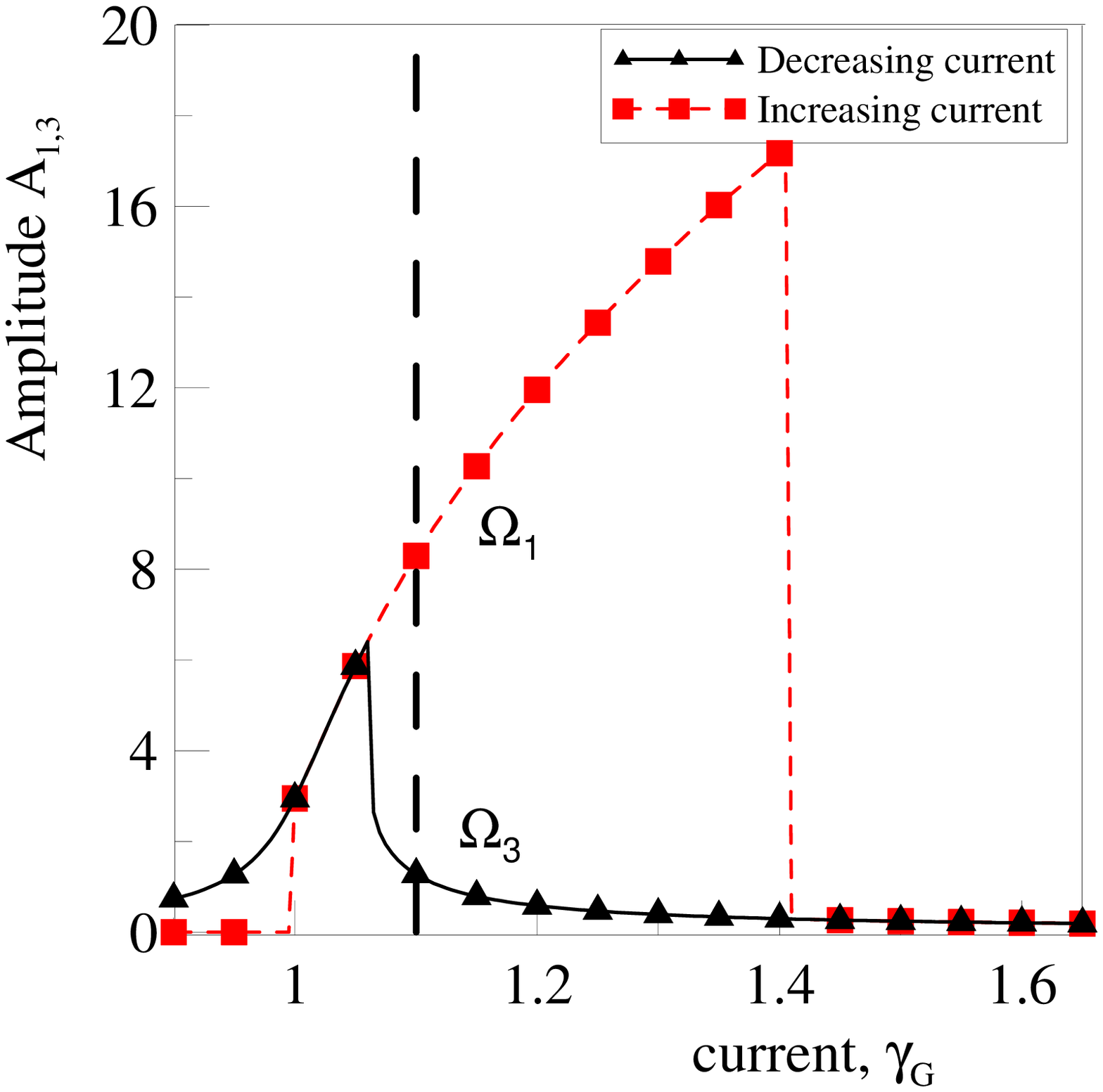}}
\centerline{ }
\centerline{ (ii)}
\centerline{\includegraphics[angle=0,scale=0.5]{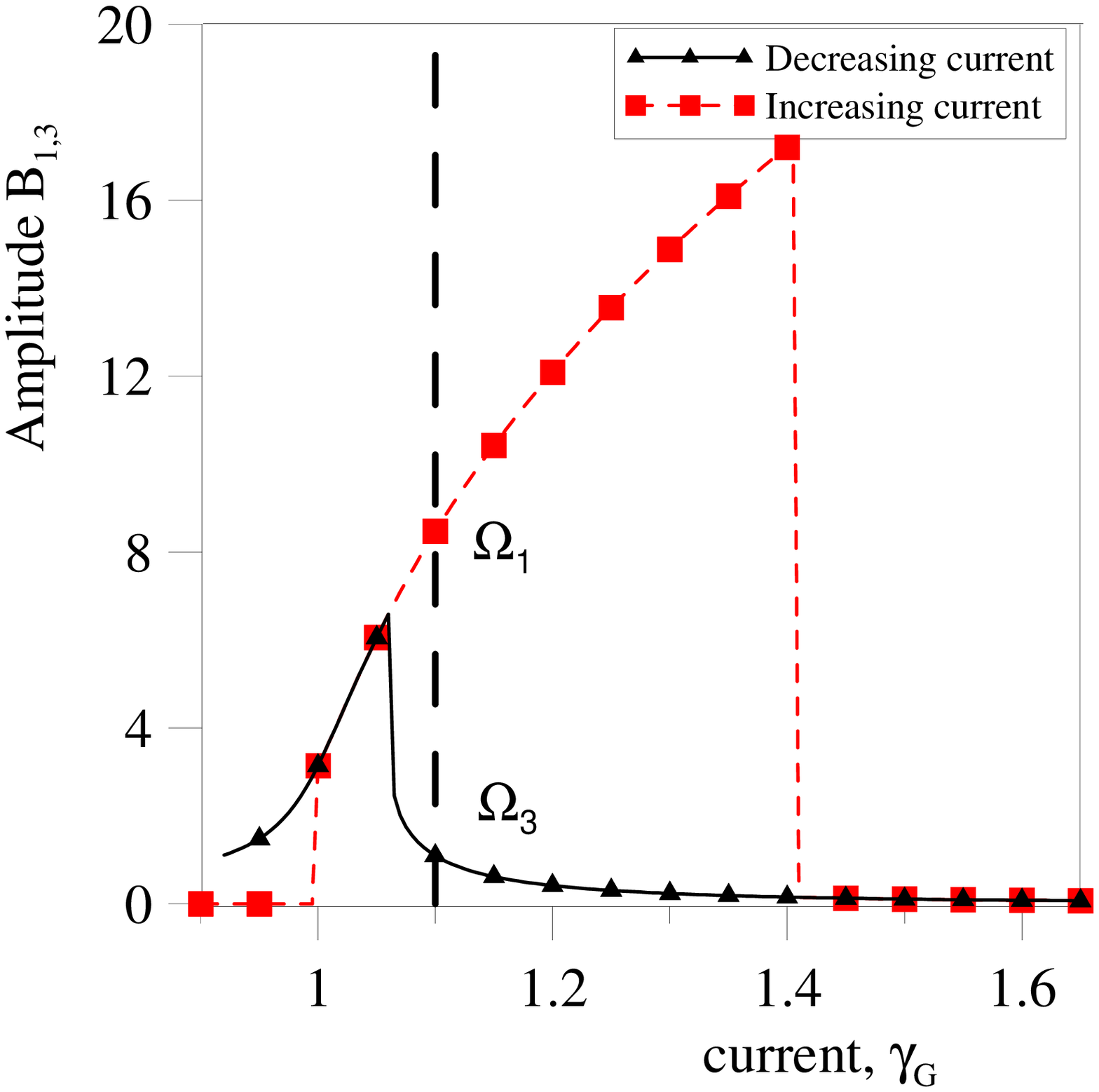}}
\caption{ \it (color online) Amplitudes $A$ and $B$ versus $\gamma_G$. Squares refer to increasing bias, and triangles to a decreasing bias, see Fig. \ref{IV}. Parameters of the simulation are: $\beta_l=0.01$, $Q_l=200$, $\Omega=2$, $\alpha=0.1$.  The long dashed line represents the bias point $\gamma_G = 1.1$ of Fig. \ref{attractor}.
}
\label{amplitude}
\end{figure}

\section{Attractors properties of birhythmic Josephson Junctions}

\begin{figure}
\centerline{\includegraphics[angle=0,height=6cm, width=10cm]{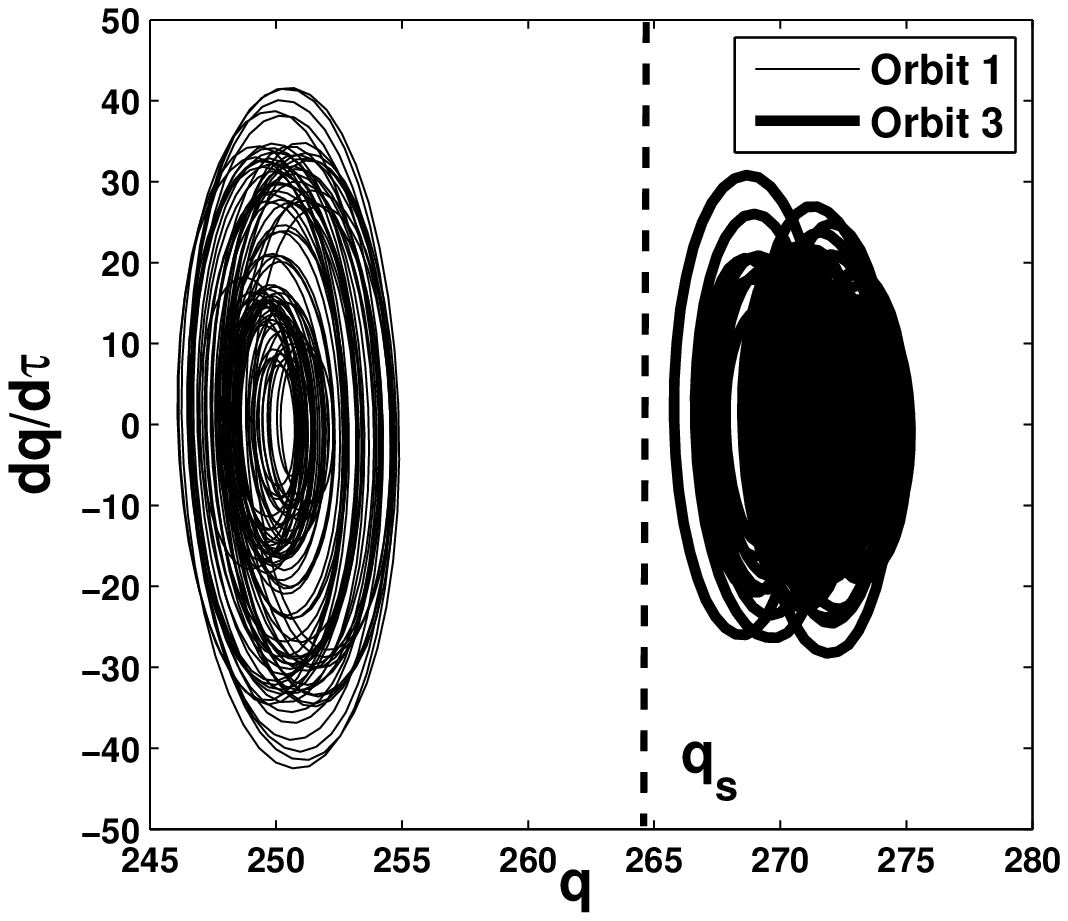}(i)}
\centerline{\includegraphics[angle=0,height=6cm, width=10cm]{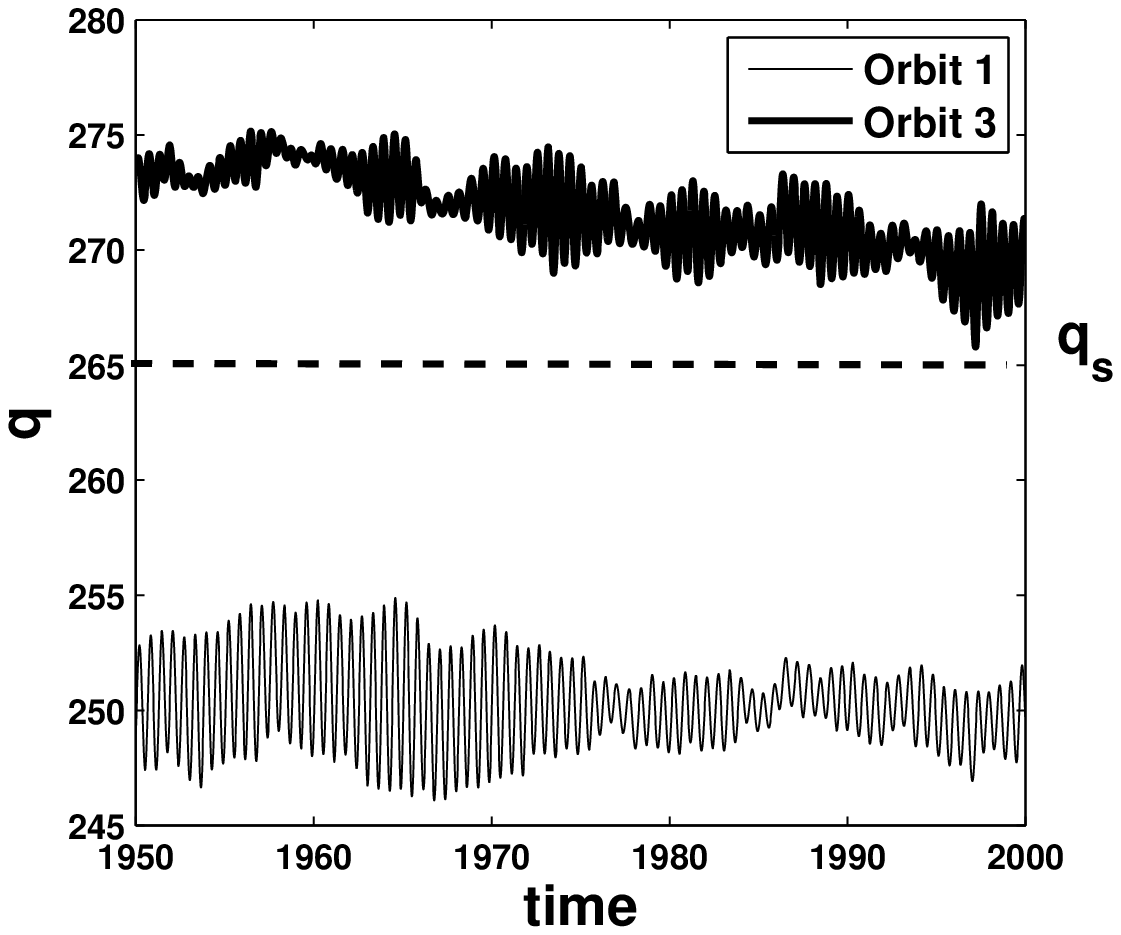}(ii)}
 \caption{\it Projection of the phase space in the $q-dq/d\tau$ plane (i) and time evolutions (ii) for the stochastic dynamics, $D=0.1$.
The parameters are the same as in Fig. \ref{attractor}.
}
 \label{attracnoise}
 \end{figure}

Figure \ref{attractor} displays the projection of the phase space in the plane $q,dq/d\tau$. The branch locked to the resonator (characterized by the frequency $\Omega_1$) -- quite naturally -- exhibits much larger excursions of the charge oscillations respect to the unlocked branch (characterized by the frequency $\Omega_3$).
In between, one postulates the existence of an unstable orbit with frequency $\Omega_2$ that represents the separatrix. 
Fig. \ref{attractor}(ii) confirms that  one can identify the attractor by the amplitude of the oscillations. In fact in Fig. \ref{amplitude} it is shown the behavior of the amplitude of the voltage  (Fig. \ref{amplitude}i ) and charge (Fig. \ref{amplitude}ii) oscillations while the bias is sweeped. The amplitudes are defined as the largest excursions of the phase derivative $d\phi/d\tau$ (proportional to the JJ voltage) and of the charge  $q$ (proportional to the capacitor voltage):

\begin{eqnarray}
\label{oscillations}
A & =& \max_{\tau} \frac{d \phi}{d\tau} - \min_{\tau} \frac{d \phi}{d\tau}  \nonumber \\
B &=&  \max_{\tau} q - \min_{\tau} q
\end{eqnarray}

\begin{figure}
\centerline{\includegraphics[angle=0,height=6cm, width=10cm]{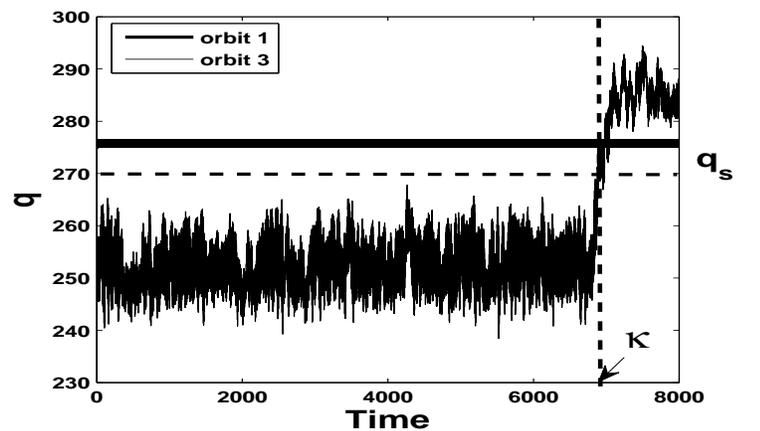}}
 \caption{\it Example of the switch from the attractor $1$ to the other under the influence of noise: after a time $\kappa \simeq 7000$ normalized units the system crosses the (estimated) separatrix $q_S=270$. Parameters of the simulation are: $\gamma_G=1.15$, $\beta_l=0.01$, $Q_l=200$, $\Omega=2$, $\alpha=0.1$.}
 \label{switch}
\end{figure}

\begin{figure}
\centerline{\includegraphics[scale=0.5]{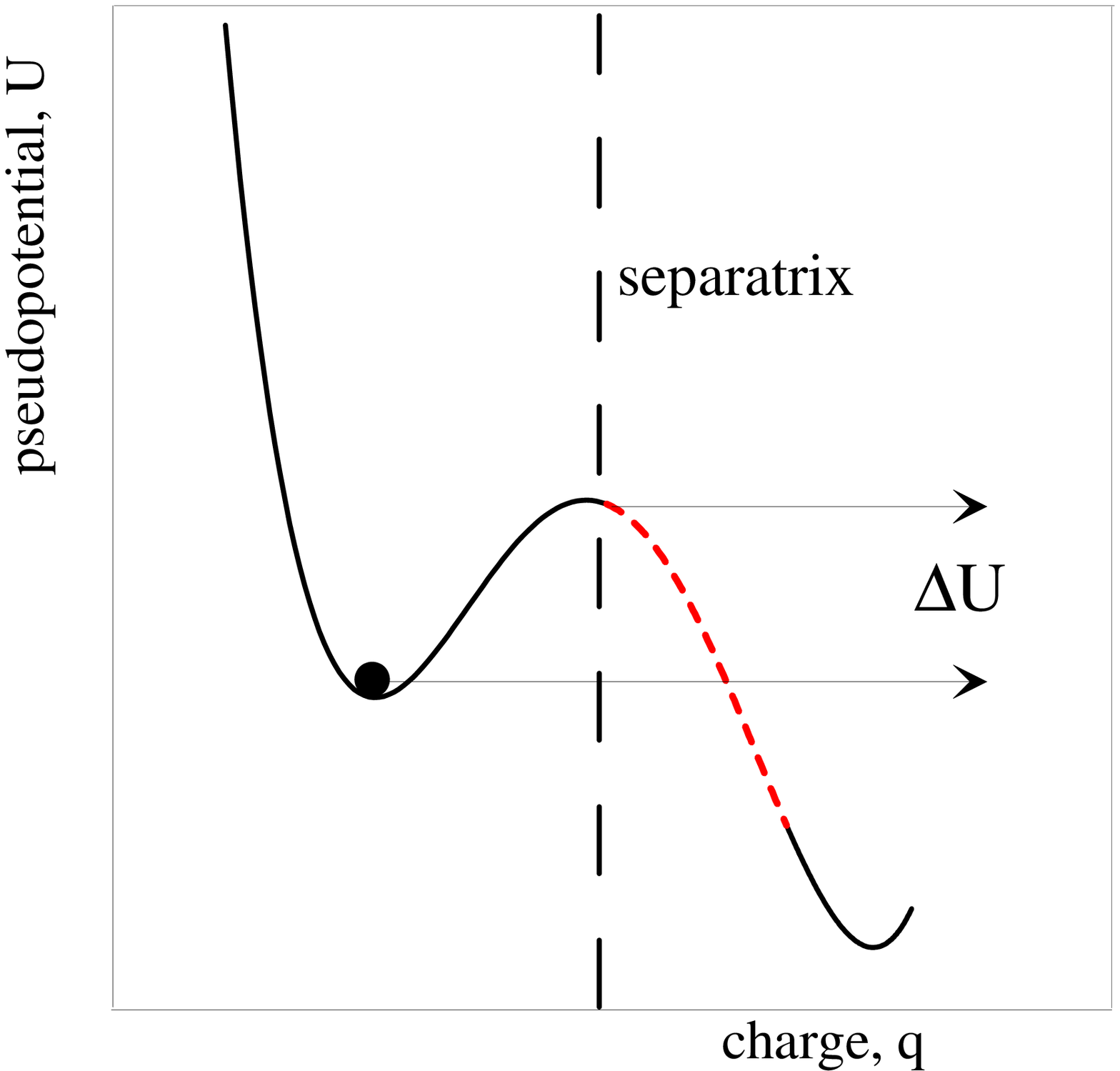}}
 \caption{\it (color online) Sketch of the escape process. The dashed part of the pseudopotential represents the zone where the threshold can be assumed without a significant change in the MFPT evaluation -- see Fig. \ref{switch}.}
 \label{potential}
\end{figure}

\noindent In Fig. \ref{amplitude} it is evident that a sudden change of the amplitudes $A$ and $B$ occurs both for low and high bias. It is exactly this sharp change that we want to exploit to retrieve the escape rate. Let us consider the dynamics under the influence of noise, Fig. \ref{attracnoise}. 
\begin{figure}
\centerline{\includegraphics[angle=0,height=5cm, width=7cm]{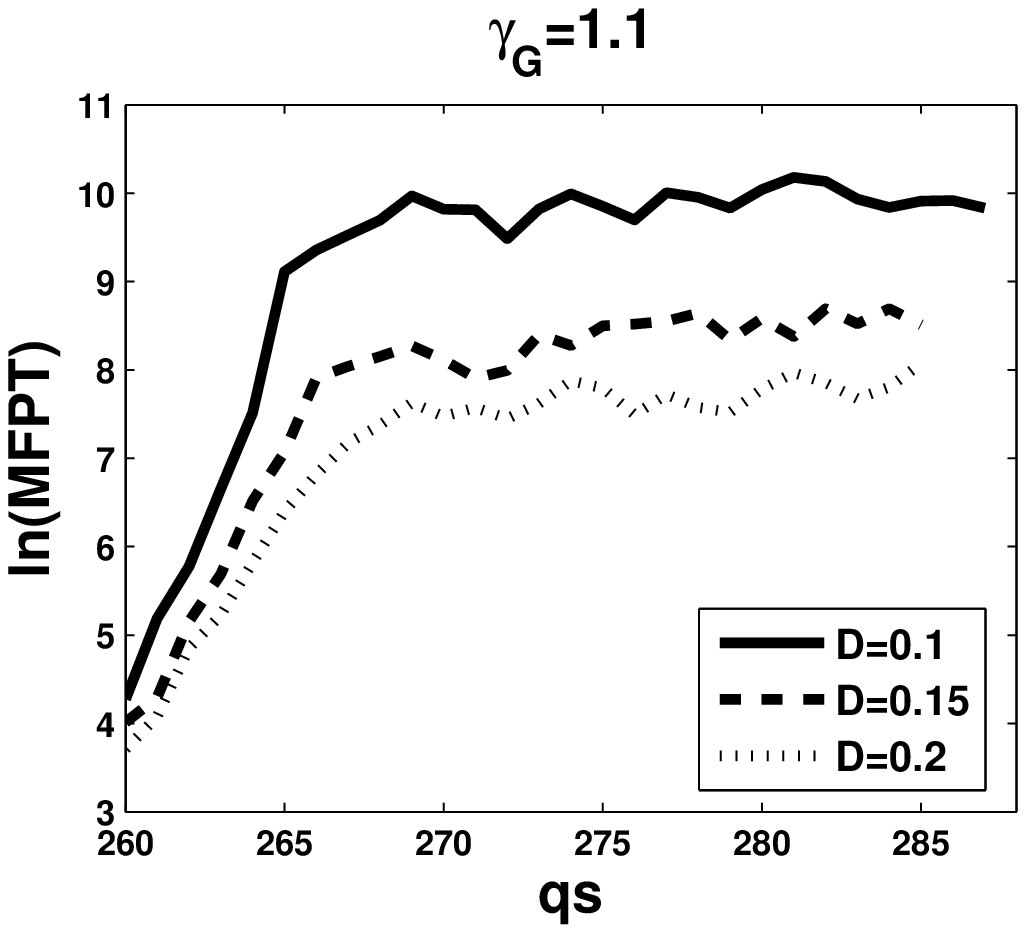}}
\centerline{\includegraphics[angle=0,height=5cm, width=7cm]{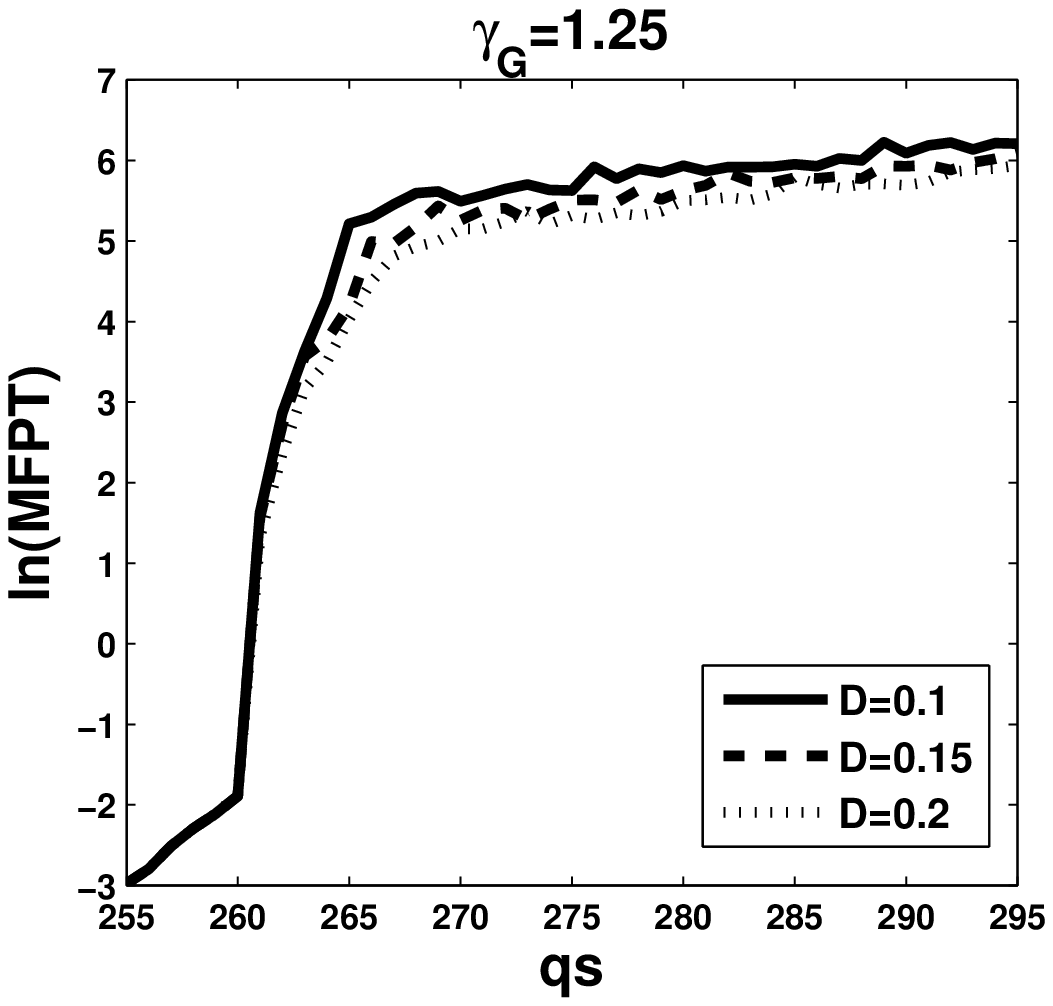}}
\centerline{\includegraphics[angle=0,height=5cm, width=7cm]{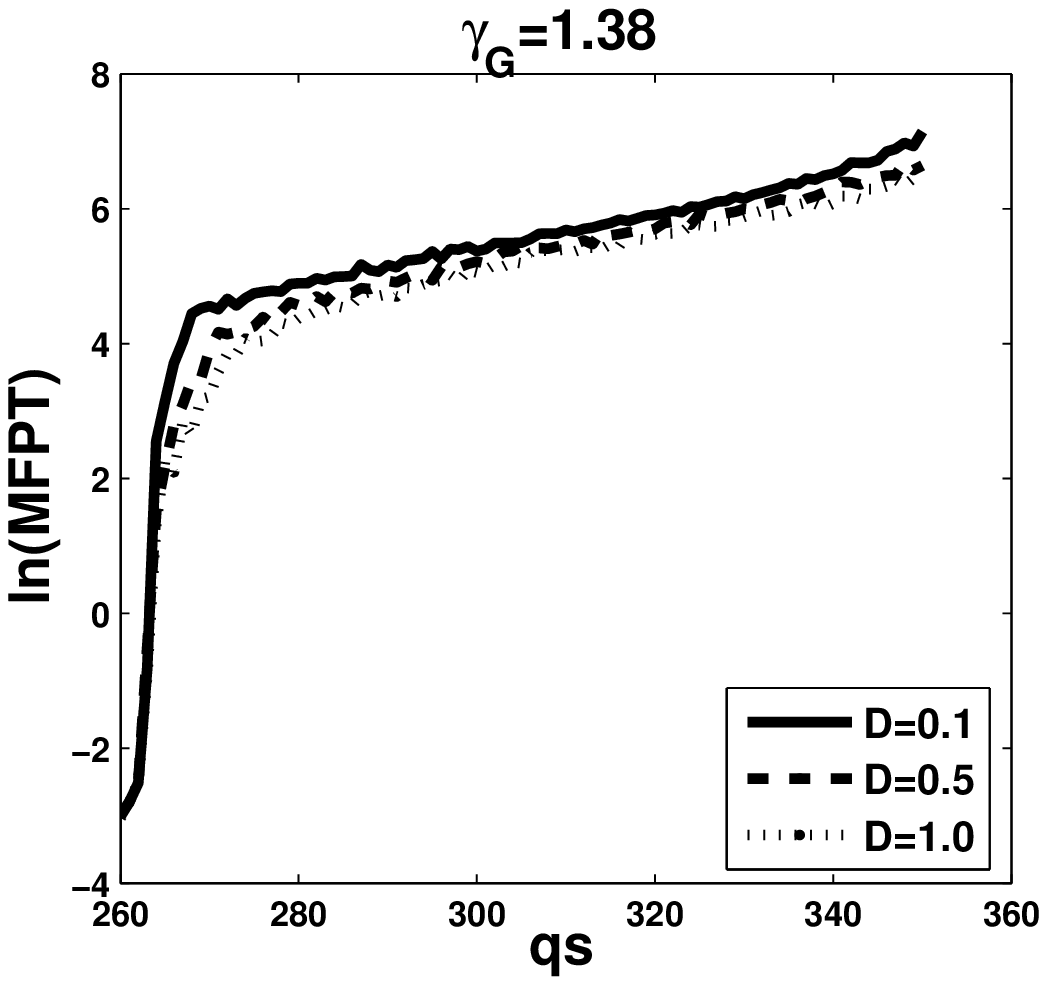}}
 \caption{\it Average MFPT  from as a function of the threshold $q_s$ at different values of  the applied current $\gamma_G$.
 Parameters of the simulation are: $\beta_l=0.01$, $Q_l=200$, $\Omega=2$, $\alpha=0.1$}
 \label{MFPT}
\end{figure}
The attractors are deformed, but still well separated, see Fig. \ref{attracnoise}i; we can therefore tentatively locate the separatrix at $q_S=265$. 
Being the system 4-dimensional the separatrix is a volume in 4-D space, whose projection in the $q,dq/d\tau$ plane ought not to be a line, and hence the dashed segment of Fig. \ref{attracnoise} is but a rough approximation. 
\begin{figure}
\centerline{\includegraphics[angle=0,height=5cm, width=7cm]{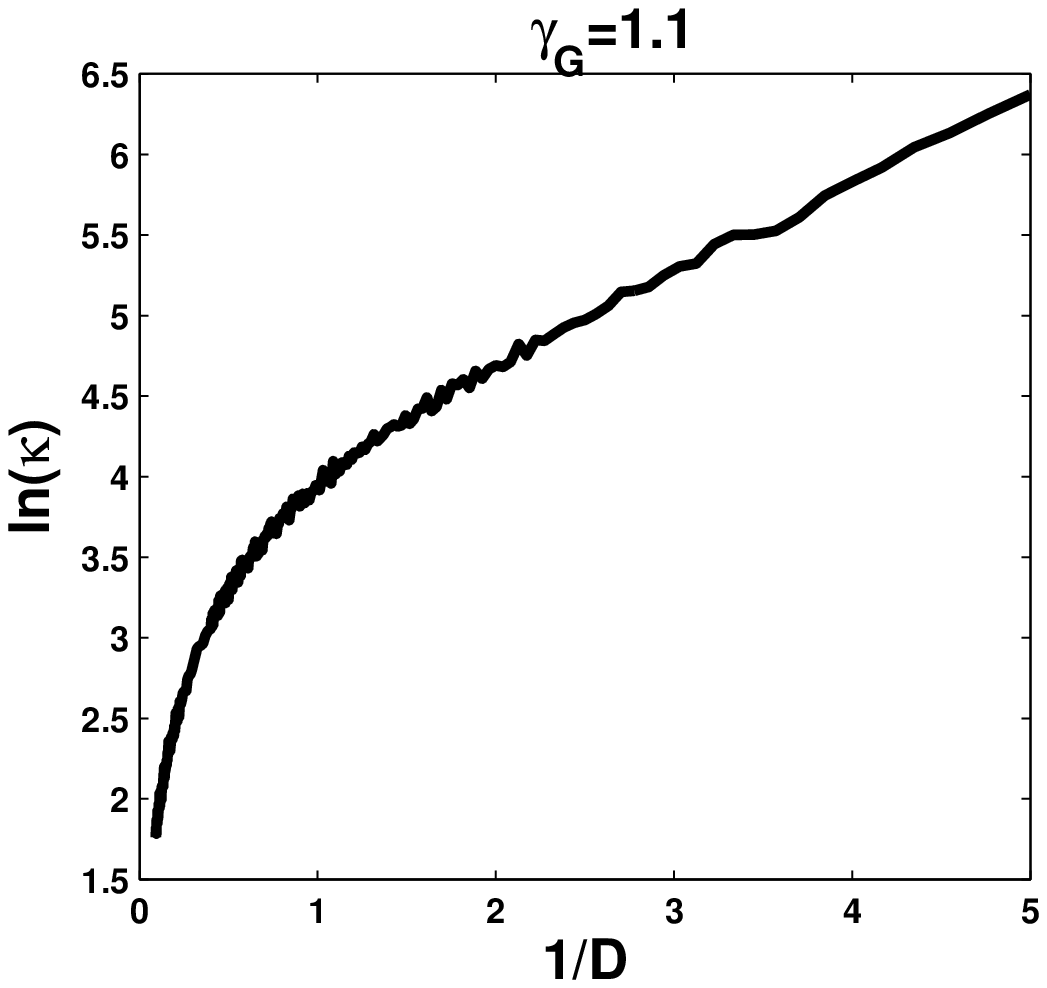}}
\centerline{\includegraphics[angle=0,height=5cm, width=7cm]{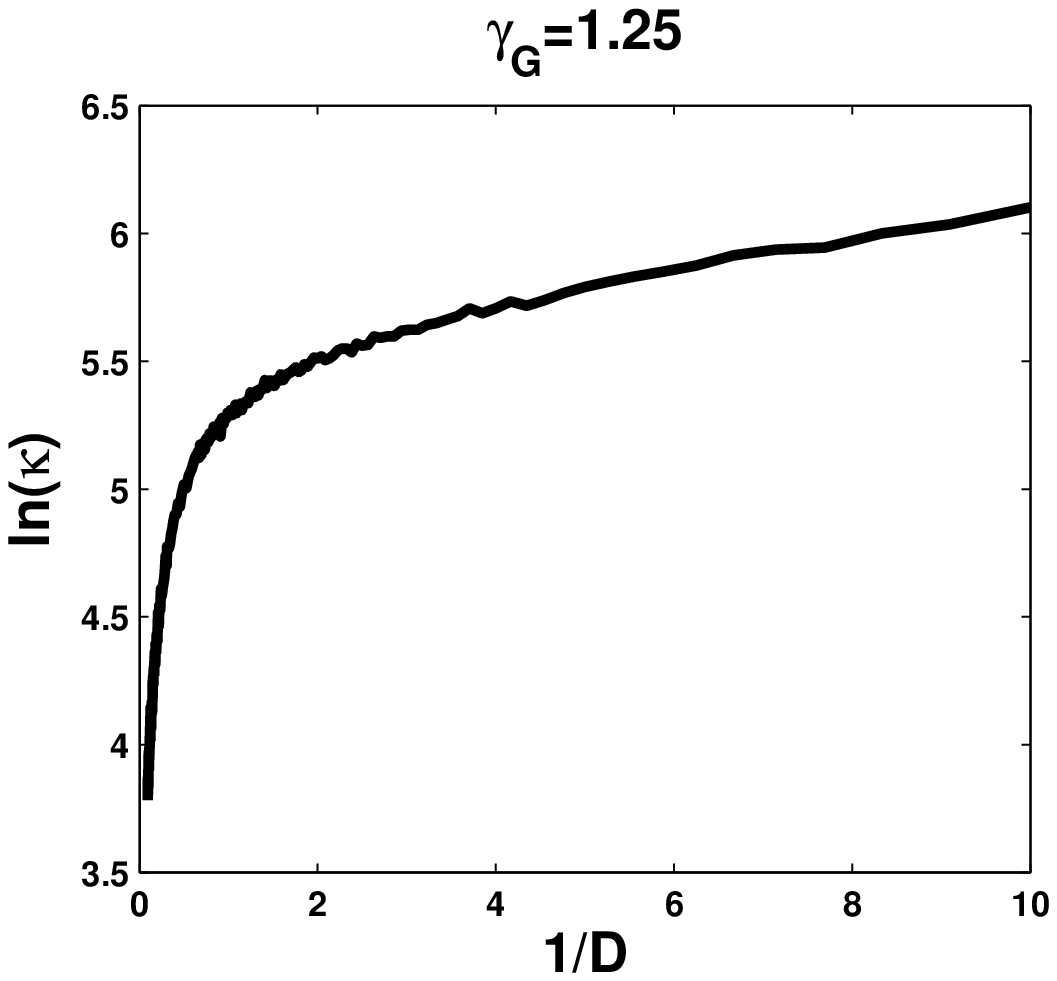}}
\centerline{\includegraphics[angle=0,height=5cm, width=7cm]{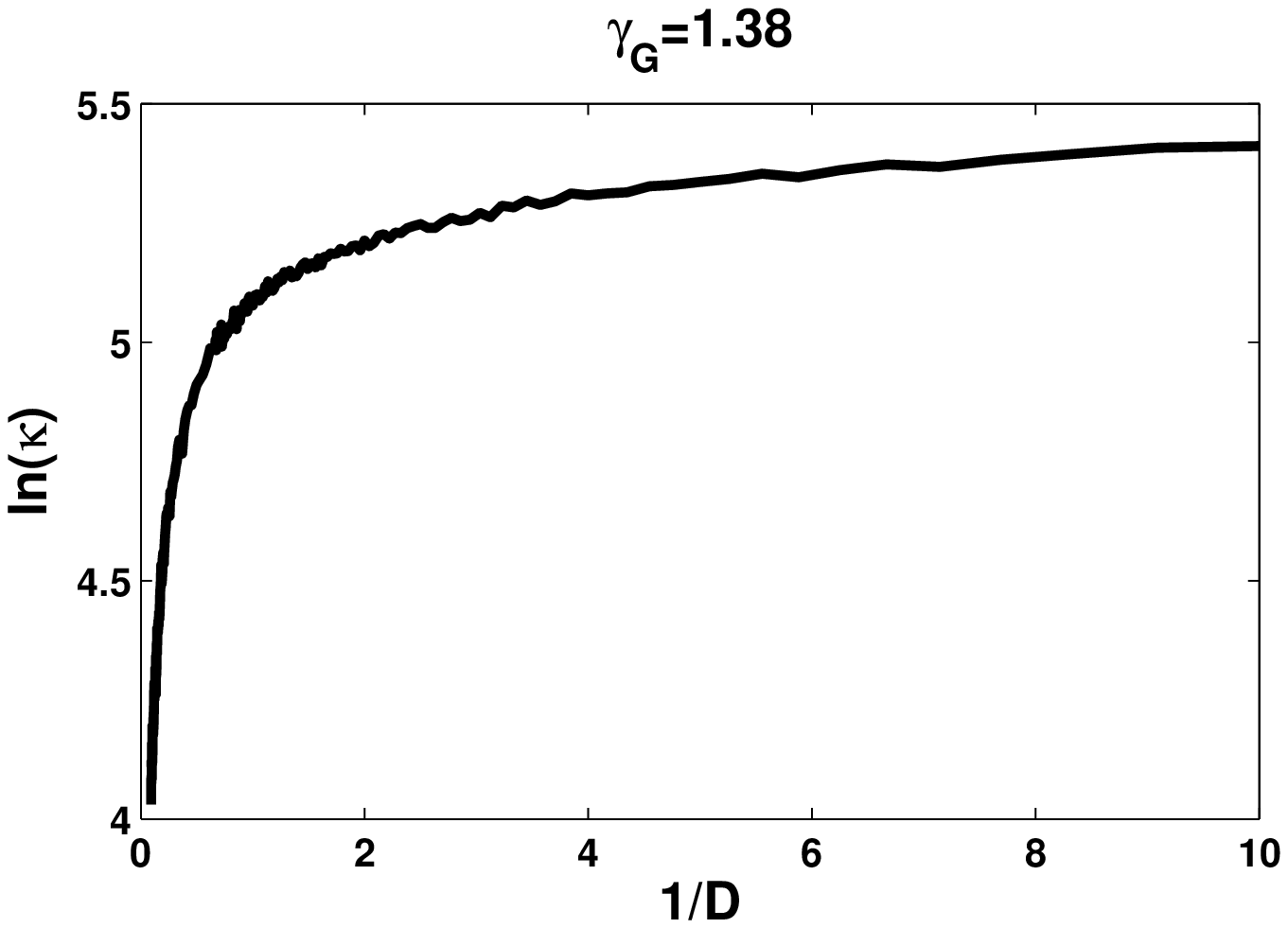}}
 \caption{\it Average of the escape time from an attractor as a function of the inverse noise intensity $1/D$ for different values of the applied current $\gamma_G$.
 Parameters of the simulation are: $\beta_l=0.01$, $Q_l=200$, $\Omega=2$, $\alpha=0.1$}
 \label{fig:kramer}
\end{figure}
We postulate that when the charge passes the threshold $q_S$ a switch occurs to the other attractor, as shown in Fig. \ref{switch}. 
In general the construction of the whole pseudopotential landscape  to identify the separatrix requires the solution of a variational problem \cite{dykman94,kautz94}.
We instead adopt a simpler procedure based on the observation that the pseudopotential is a Lyapunov function \cite{graham85}, and therefore becomes negative beyond the separatrix, as schematically shown in Fig. \ref{potential}. 
{
In fact Fig. \ref{switch} illustrates that a sudden switch occurs when the fluctuations exceed a threshold, or when $q>q_s$, i.e. when the system passes into the descending part of the pseudopotential in Fig. \ref{potential}. 
However, exactly because of the switch towards the attractor, it is not necessary to accurately know the value of $q_s$: any value behind $q_s$ leads to a similar estimate of the MFPT (Mean First Passage Time) \cite{hanggi90}, see Fig. \ref{MFPT}. 
We emphasize that the mean first passage time across any point in the vicinity of the separatrix has two distinct behaviors: i) it increases exponentially when the threshold point $q_s$ is set before the separatrix, and ii) it increases very weakly when the threshold $q_s$ is beyond the maximum of the potential. The different behavior is shown in Fig. \ref{MFPT}, and therefore from the change in the slope of the MFPT we estimate the position of the separatrix.
}
In summary, in the descending region beyond the separatrix (the dashed part of the pseudopotential) the system quickly runs "downhill", and the time elapsed in the dashed part is negligible respect to the time necessary to reach, under the influence of noise, the peak of the pseudopotential. 
This conjecture is confirmed by the MFPT with different choices of the threshold $q_S$ (see Fig. \ref{MFPT}): there is a region where the average time is almost independent of the choice of the threshold. We conclude that the knee of the MFPT can be used as an effective separatrix to estimate the pseudoenergy activation barrier. 

This is practically implemented in Fig. \ref{fig:kramer} for different values of the bias current. 
The linear relationship between the logarithm of the escape time and the inverse of the noise intensity offers the estimate of an effective energy barrier, see Eq.(\ref{pseudopot}):

\begin{eqnarray}
\label{kramer}
\Delta U \simeq \frac{\Delta ln(\kappa)}{\Delta \left( 1/D \right)}
\end{eqnarray}

\noindent Equation (\ref{kramer}) is the main result of this part of the paper: to characterize with an activation energy the metastable states in the birhythmic region. 

\section{Energy barrier and lifetime of the $RLC$ induced step}

In this Section we collect the results on the analysis of the birhythmic region of the IV curve in Fig. \ref{IV}. 

\begin{figure}
\centerline{\includegraphics[scale=0.5]{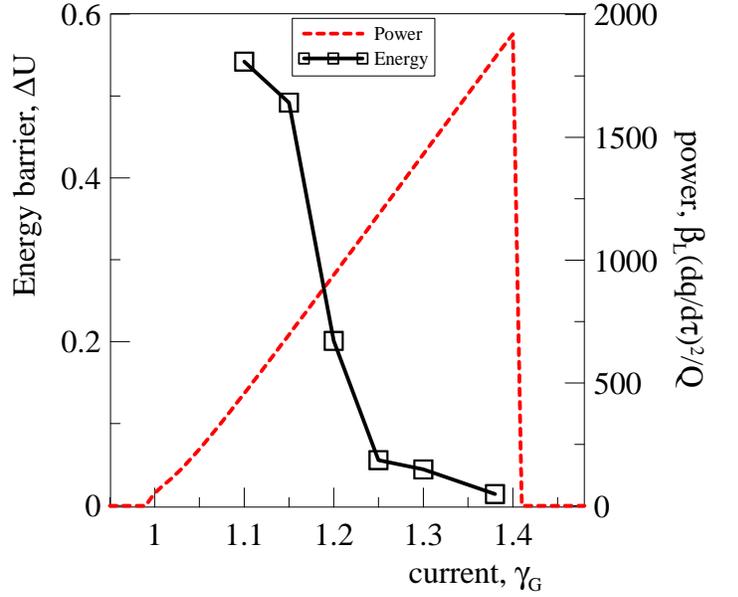}}
 \caption{\it (color online) Energy barrier to escape from the attractor $1$ and power dissipated in the load as a function of the applied bias current. Parameters of the simulation are: $\beta_l=0.01$, $Q_l=200$, $\Omega=2$, $\alpha=0.1$, as in Fig. \ref{amplitude}.}
\label{energy}
\end{figure}
We use Eq.(\ref{kramer}) to retrieve the behavior of the activation energy as a function of the bias $\gamma_G$, see Fig. \ref{energy}. The energy barrier for low bias $\gamma_G$ is large and the attractor is bounded in a stable well. When the current is increased along the step the energy barrier decreases, and almost disappears at the top of the step, where the frequency splitting is at a maximum, see Fig. \ref{omega}. 
In the same process, the energy dissipated by the cavity increases, for the normalized power linearly rises along the step: 

\begin{eqnarray}
\label{power}
\frac{P}{V_NI_0} = \frac{1}{V_NI_0} R \frac{1}{T} \int_0^T \left(\frac{d\tilde q}{d t}\right)^2 dt \nonumber \\
= \frac{\beta_L}{Q} \frac{1}{\omega_j T} \int_0^{\omega_j T} \left(\frac{d q}{d \tau}\right)^2 d\tau 
= \frac{\beta_L}{Q} <\left(\frac{dq}{d\tau}\right)^2>
\end{eqnarray}

\noindent  (in the dimensionless equations we are using the voltage is expressed in $V_N = (\hbar I_0/2eC_j)^{1/2}$ units). At the bottom of the step ($\gamma_G \simeq 1.05$) the power is low, while the energy barrier is at the maximum. In the region  $1.15 \le \gamma_G \le 1.25$ the effective energy barrier decreases of about an order of magnitude ($\Delta U$ passes from $\Delta U \simeq 0.54$ to $\Delta U \simeq 0.055$). 
{
A relevant feature of Fig. \ref{energy} is that the change of the pseudopotential in the birhythmic region cannot be ascribed to a difference in the frequencies. In fact the pseudopotential is at a maximum when the difference is at a minimum. Thus, the switch from the resonant step back to the IV curve of the unperturbed dynamics occurs because the activation pseudopotential vanishes.  
}
The change is dramatic if one considers that time is normalized respect to $\omega_j$, that is typically above 100GHz. 
A lifetime of the order of a second therefore entails a noise level $D$ as low as to reach $\kappa \simeq 10^{11}$, or $ln(\kappa)\simeq 25$. 
From the behavior shown in Fig. \ref{fig:kramer} one estimates $D \simeq 0.025$ for $\gamma_G=1.15$, and $D \simeq 0.0028$ for $\gamma_G=1.25$. 
Put it another way, at a fixed noise level $D=0.25$ the lifetime decreases of at least seven decades when the bias current passes from $\gamma_G=1.15$ to $\gamma_G=1.25$.
It is also noticeworthy that the Arrhenius-like behavior implied by the existence of the pseudopotential greatly simplifies the numerical problem to find the noise intensity at which the desired lifetime is reached. 
In fact the relation Eq.(\ref{pseudopot}) allows to extrapolate very long lifetimes from Fig. \ref{fig:kramer}, whereas direct simulations of such long lifetimes are prohibitive.

\section{Conclusions}
We have found that the global stability analysis of JJ coupled to a resonator shows a striking change in the birhythmic region: the attractor characterized by a frequency locked to the resonator is most stable for low bias current, when the power dissipated in the cavity is small. The system is, unfortunately, less stable at the top of the step, when the current in the resonator is at a maximum and the two frequencies are most separated ( a similar conclusion has been reached for another nonlinear birhythmic system \cite{yamapi10}). Thus the analysis of large excursions, as large to drive the system from the desired attractor to another, indicates that the global stability is weak where most power is available. This observation is, in some sense, bad news for applications, inasmuch it shows that stability and high power are contradictory requirements. However, the detailed behavior of the global stability demonstrates that the deterioration occurs at the middle of the step, where the stability is still relatively high -- see Fig. \ref{energy}. 
{
From a more general point of view, we have shown that the stability of a dynamic state can be analyzed in terms of the pseudopotential also when the separatrix is not known and the variational approach is difficult to apply. Instead we propose a simpler method to (approximately) determine the position of the separatrix from the change in the slope of the MFPT. 
}

A number of cautions are in order, however. In the first place, we have analyzed a single junction coupled to a cavity, while for applications such as BSCCO stacks one should consider many junctions \cite{tachiki11,welp13}. 
Second, we are using lumped elements for both the junctions and the cavity, whereas a distributed description \cite{tachiki11,welp13,sakai93,madsen04,madsen08} is more appropriated. 
Finally, thermal effects cause self-heating and back-bending of the $IV$ usually associated to $THz$ emission \cite{ozyuzer07,gross13}, whereas we have here only addressed the effect of random fluctuations. Nevertheless, the calculations of this work point to a conceivable danger: that global stability properties are of crucial importance to determine the region of parameters where large power devices could possibly work.

\section*{Acknowledgements}
R.Yamapi undertook this work with the support of the ICTP (International Centre for Theoretical Physics) in the framework of Training and Research in Italian Laboratories (TRIL) for AFRICA programme, Trieste, Italy. He also acknowledges the hospitality of the Dipartimento di
Fisica "E.R. Caianiello" of the Universit\`a di Salerno, Fisciano, Italy. 

The authors acknowledge partial financial support from PON Ricerca e Competitivit\`a 2007-2013 under grant agreement PON NAFASSY, PONa3\_00007.

\newpage


\end{document}